\documentclass[preprint,review,12pt]{elsarticle}

\usepackage{amsmath,amscd,amssymb}
\usepackage{bm}
\usepackage{graphicx}
\usepackage[tight]{subfigure}
\usepackage{units}
\usepackage{array,booktabs}
\usepackage{amssymb}
\usepackage{setspace}
\usepackage{lineno}
\usepackage{color}

\journal{Ultrasonics}

\begin{document}

\begin{frontmatter}
\title{Difference-frequency generation in nonlinear scattering of acoustic waves by a rigid sphere}
\author{Glauber T.\ Silva$^*$}\author{Anderson Bandeira}
\cortext[cor1]{Corresponding author: \texttt{glauber@pq.cnpq.br}}
\address{Physical Acoustics Group, Instituto de F\'isica, 
Universidade Federal de Alagoas, Macei\'o, AL 57072-970, Brazil.}

\begin{abstract}
In this paper, the partial-wave expansion method is applied to describe the difference-frequency pressure generated 
in a nonlinear scattering of two acoustic waves with an arbitrary wavefront by means of a rigid sphere. 
Particularly, the difference-frequency generation is analyzed in the nonlinear scattering with a spherical 
scatterer involving two intersecting plane waves in the following configurations: 
collinear, crossing at right angles, and counter-propagating.
For the sake simplicity, the plane waves are assumed to be spatially located in a spherical 
region which diameter is smaller than the difference-frequency wavelength.
Such arrangements can be experimentally accomplished in vibro-acoustography 
and nonlinear acoustic tomography techniques.
It turns out to be that when the sphere radius is of the order of the primary wavelengths,
and the downshift ratio (i.e. the ratio between the fundamental frequency and the difference-frequency) is larger than five,
difference-frequency generation is mostly due to a nonlinear interaction between the primary scattered waves. 
The exception to this is the collinear scattering for which the nonlinear interaction of the primary incident 
waves is also relevant. 
In addition, the difference-frequency scattered pressure in all scattering configurations decays 
as $r^{-1} \ln r$ and $1/r$, whereas $r$ is the radial distance from the scatterer to the observation point.
\end{abstract}

\end{frontmatter}

\linenumbers

\section{Introduction} 
An outstanding feature of the nonlinear interaction of two or more acoustic waves is a generation 
of secondary waves having different frequencies, namely
harmonics, sum- and difference-frequency~\cite{thuras:173}.
In the presence of an inclusion, this generation is enhanced by two physical effects. 
First of all, the incident waves produce a radiation force through nonlinear interactions with the 
inclusion~\cite{silva:056617,silva:026609,
silva:234301,silva:48}. 
As a result, the inclusion  is set in motion emitting waves which frequencies correspond to 
the components present in the dynamic radiation force. 
In addition, the primary waves (incident and scattered) related to the fundamental 
frequencies interact yielding secondary waves. 
This process is also known as scattering of sound-by-sound in which sum- and difference-frequency waves are 
generated~\cite{ingard:367, westervelt:199, jones:398,berntsen:1968}.

Difference-frequency generation is present in several applications like parametric 
array sonar~\cite{mellen:932}, 
audio spotlight~\cite{yoneyama:1532}, characterization of liquid-vapor phase-transition~\cite{mujica:234301}, 
and acoustical imaging methods such as nonlinear parameter 
tomography~\cite{sato:295,sato:49,kim:175, zhang:1359} and 
vibro-acoustography~\cite{fatemi98a,urban:1284,silva:5985}.
Moreover, parametric arrays have been used to produce low-frequency waves in wideband scattering 
experiments~\cite{humphrey:265}.
In this case, the scatterer is placed outside the interaction region of the incident waves
and the scattering  is treated through the linear scattering theory.
This is similar to calibrating parametric sonars based
on measurements of the linear scattering cross-section~\cite{foote:1482}.

Investigations of difference- and sum-frequency generation concerning to spherical and cylindrical scattered
waves were 
firstly performed by Dean-III~\cite{dean:1039}.
Scattering consisting of  nonlinear interaction of a plane wave with a radially vibrating rigid 
cylinder~\cite{piquette:880} and sphere~\cite{lyamshev:50} have also been analyzed.
Moreover, difference-frequency generation in  scattering of 
two collinear plane waves by means of a sphere was previously studied~\cite{abbasov:158}.
However,  the results obtained in this study show that the difference-frequency scattered pressure 
has singularities in the polar angle of spherical coordinates
(i.e. the angle formed by the position vector and the $z$-axis).
Furthermore, the difference-frequency scattered pressure  only depends on the monopole terms of the 
primary waves.
Giving this physical picture, a broader discussion is required on how to handle the singularities
and why the information from  higher-order multipole terms of the primary waves were discarded.

Applications of difference-frequency generation in acoustics generally employ incident beams 
which deviate from collinear plane waves.
This has stimulated the investigation of nonlinear scattering of 
two acoustic waves with an arbitrary wavefront.
Our analysis stems from the Westervelt wave equation~\cite{westervelt:535}.
This equation is solved through the method of successive approximations in addition to 
the Green's function technique.
Furthermore, appropriate boundary conditions are established to garantee a unique solution of the
Westervelt equation.
The difference-frequency scattered pressure is obtained as a partial-wave expansion which depends 
on beam-shape and scattering coefficients.
Each of these coefficients is related, respectively, 
to a complex amplitude of a partial-wave that composes the primary
incident and scattered waves~\cite{silva:298, mitri:392}.

The method proposed here is applied to the nonlinear scattering of two intersecting plane 
waves by a rigid sphere.
The difference-frequency scattered pressure is obtained in the farfield in 
three incident wave configurations: collinear, perpendicular, and counter-propagating.
In this analysis, the downshift ratio is larger than five.
It is worthy to mention that the collinear configuration of
 incident waves has been implemented in vibro-acoustography 
experiments~\cite{fatemi98a}, while the perpendicular and  counter-propagating arrangements have been 
experimentally studied in Refs.~\cite{sato:49,kim:175}, respectively.
To reduce the mathematical complexity of the model, the incident waves are assumed to be spatially 
located in a spherical region. 
Even though this approach is not entirely realistic, experimental accomplishment of scattering 
 of two located intersecting ultrasound beams was reported in Ref.~\cite{chen:313}.

The results show that in the collinear case, the nonlinear interaction involving the primary incident waves 
(incident-with-incident interaction) and that of the primary scattered waves 
(scattered-with-scattered interaction) are responsible for difference-frequency generation. 
In the perpendicular and counter-propagating configurations,  difference-frequency generation
is mostly due to the scattered-with-scattered interaction.
In addition, the difference-frequency scattered pressure increases with difference-frequency and
varies with the  radial distance $r$ from the scatterer to observation point as $r^{-1}\ln r$ and 
$1/r$. 
A similar result is found in Ref.~\cite{dean:1039}, though only monopole sources were considered.

\section{Physical model}
\label{sec:scattering}
Consider a nonviscous fluid with an ambient density $\rho_0$ and an adiabatic speed of sound $c_0$.
The fluid is assumed to have infinite extent.
Acoustic waves in the fluid can be described by the acoustic pressure
$p$ as a function of the position vector $\bf r$ and  time $t$.
Absorption effects of a viscous fluid can be readily included for longitudinal acoustic waves 
(compressional waves).
In this case, the wavenumber of a single-frequency wave becomes a complex number.
However, the account for shear wave propagation, which is supported in viscous fluids, 
lies beyond the scope of this study.

\subsection{Wave dynamics}
We are interested on describing how a difference-frequency wave is generated in 
a nonlinear scattering of two  incident acoustic waves by means of a rigid sphere.
The scope of this analysis is limited to acoustic pressures propagating in the farfield.
Up to second-order approximation, the farfield pressure satisfies the lossless Westervelt wave equation~\cite{hamilton:3}
\begin{equation}
\label{westervelt}
 \left(\nabla^2 - \frac{1}{c_0^2}\frac{\partial^2}{\partial t^2} \right) p  = -
\frac{\beta}{\rho_0 c_0^4}\frac{\partial^2 p^2}{\partial t^2} ,
\end{equation}
where $\beta = 1+(1/2)(B/A)$, with $B/A$ being the thermodynamic nonlinear parameter of the fluid.
This equation accounts for wave diffraction and medium nonlinearity.
It is worthy to notice that Eq.~(\ref{westervelt}) is valid when cumulative effects 
(such as wave distortion) are dominant over nonlinear local effects.
This happens when the propagating wave is far from acoustic sources.
When the wave is observed near to a scatterer, its pressure should 
be modified to~\cite{hamilton:p54}
\begin{equation}
\label{pmod}
 \tilde{p} = p + \frac{\rho_0}{4} \left(\nabla^2 + \frac{1}{c_0^2}\frac{\partial^2}{\partial t^2} \right)\phi^2.
\end{equation}
where $\phi$ is the velocity potential.  
Note that the approximation $\tilde{p}=p$ holds for farfield waves.

Let us assume that the  acoustic pressure is given in terms of the Mach number 
$\varepsilon=v_0/c_0$ and $ \varepsilon \ll 1$ (weak-amplitude waves),
where $v_0$ is the maximum magnitude of the particle velocity in the medium.
Hence, we can expand the pressure up to  second-order as~\cite{hamilton:281}
\begin{equation}
\label{p_expansion}
 p =  \varepsilon p^{(1)} + \varepsilon^2p^{(2)},\quad \varepsilon \ll 1
\end{equation}
where  $p^{(1)}$, and $p^{(2)}$ are, respectively,  the linear (primary) and  the second-order (secondary) pressure fields.
In the weak-amplitude approximation ($\varepsilon\ll 1$), the primary
and the secondary pressures suffice to describe nonlinear effects in wave propagation.
Now, substituting Eq.~(\ref{p_expansion}) into Eq.~(\ref{westervelt}) and grouping terms of like powers $\varepsilon$ 
and $\varepsilon^2$, one obtains
\begin{align}
\label{primary}
 \left(\nabla^2 - \frac{1}{c_0^2}\frac{\partial^2 }{\partial t^2} \right)p^{(1)} &= 0,\\
\left(\nabla^2 - \frac{1}{c_0^2}\frac{\partial^2 }{\partial t^2} \right)p^{(2)} &= 
-\frac{\beta}{\rho_0 c_0^4}\frac{\partial^2 p^{(1)2}}{\partial t^2}.
\label{secondary}
\end{align}
These equations form a set of hierarchical linear wave equations.

\subsection{Linear scattering}
\begin{figure}[h]
\centering
\includegraphics[width=.3\textwidth]{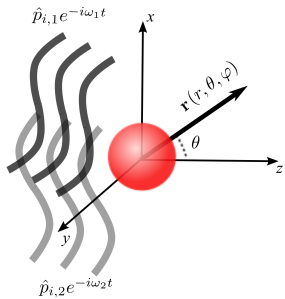}
\caption{\label{fig:system} (Color online) Outline of the scattering problem.
Two incident waves of arbitrary wavefront with amplitudes $\hat{p}_{i,1}$ and $\hat{p}_{i,2}$
insonify a target.
The observation point is denoted in spherical coordinates by ${\bf r}(r,\theta,\varphi)$, 
where $r$ is the radial distance from the scatterer to the observation point, $\theta$ and $\varphi$ are the polar and the azimuthal angles, respectively.}
\end{figure}

Assume that two primary acoustic waves of 
arbitrary wavefront with frequencies $\omega_1$ and $\omega_2$ ($\omega_2>\omega_1$), propagate toward a scatterer 
suspended in a host fluid. 
The total incident pressure due to the waves is given by
\begin{equation}
\label{pi}
p_i = \varepsilon \rho_0 c_0^2 (\hat{p}_{i,1}e^{-i\omega_1t}  +\hat{p}_{i,2}e^{-i\omega_2 t}),
\end{equation}
where $i$ is the imaginary unit, $\hat{p}_{i,1}$ and $\hat{p}_{i,2}$ are the dimensionless pressure 
amplitudes of the incident waves.
When the scatterer is placed in the interaction region 
of the incident waves (see Fig.~\ref{fig:system}), two primary scattered waves appear in the medium.
Hence, the primary scattered pressure reads
\begin{equation}
\label{ps1}
p_s = \varepsilon \rho_0 c_0^2 (\hat{p}_{s,1}e^{-i\omega_1t}  +\hat{p}_{s,2}e^{-i\omega_2 t}),
\end{equation}
where $\hat{p}_{s,1}$ and $\hat{p}_{s,2}$ are the dimensionless pressure amplitudes of the scattered waves.
Therefore, the total primary pressure in the fluid is then $p^{(1)}=p_i + p_s$.

It is worthy to notice that the quadratic term $\partial^2 p^{(1)2}/\partial t^2$ in Eq.~(\ref{secondary}) 
gives rise to waves at  second-harmonic frequencies $2\omega_1$ and $2\omega_2$, sum-frequency $\omega_1+\omega_2$, 
and difference-frequency 
$\omega_2-\omega_1$. 
These frequency components are distinct and do not affect each other.
Our analysis is restricted to difference-frequency component only.

By substituting Eqs.~(\ref{pi}) and (\ref{ps1}) into Eq.~(\ref{primary}), we find that the primary pressure amplitudes
satisfy the Helmholtz equation
\begin{equation}
 \left(\nabla^2  + k_n^2\right) \left(
\begin{matrix}
\hat{p}_{i,n}\\
\hat{p}_{s,n}
\end{matrix}
\right) = 0,  \quad n=1,2,
\label{helm1}
\end{equation}
 where $k_n=\omega_n/c_0$ is the primary wavenumber.

The incident pressure amplitudes are assumed to be regular (finite) in the origin of the coordinate system. 
Thus, they are given, in  spherical coordinates (radial distance $r$,  
polar angle $\theta$,  azimuthal angle $\varphi$) by~\cite{williams:6}
\begin{equation}
\label{p_i}
\hat{p}_{i,n} =\sum_{l,m} a_{nl}^m j_l(k_n r)Y_l^m(\theta,\varphi), \quad n=1,2,
\end{equation}
where $\sum_{lm} \rightarrow \sum_{l =0}^{\infty}\sum_{m=-l}^l $, $a_{nl}^m$ is the beam-shape coefficient,  $j_l$ is the 
spherical Bessel function of $l$th-order and $Y_l^m$ is the spherical harmonic of $l$th-order 
and $m$th-degree.
The beam-shape coefficients can be determined by using the orthogonality property
of the spherical harmonics.
Numerical quadrature can be used to compute these coefficients for waves with arbitrary wavefront~\cite{silva:298,mitri:392}.

The scattered pressure amplitudes are given by~\cite{williams:p206}                                                                                         
\begin{equation}
 \label{scattered}
 \hat{p}_{s,n} = \sum_{l,m} s_{nl}^m h^{(1)}_l(k_n r) Y_l^m(\theta,\varphi), \quad n=1,2 
\end{equation}
where $h^{(1)}_l$ is the first-type spherical Hankel function of $l$th-order 
and $s_{nl}^m$ is the scattering coefficient
to be determined from acoustic boundary conditions on the scatterer's surface.

\subsection{Difference-frequency generation}
The generated difference-frequency pressure is a second-order field in
the Mach number expansion~(\ref{p_expansion}).
Thus, we may express the difference-frequency pressure as
\begin{equation}
\label{pm_time}
 p_- = \varepsilon^2 \rho_0 c_0^2 \hat{p}_- e^{-i \omega_- t}, 
\end{equation}
where $\hat{p}_-$ is the dimensionless difference-frequency pressure amplitude and 
$\omega_-=\omega_2 - \omega_1$.
Substituting Eq.~(\ref{pm_time}) into Eq.~(\ref{secondary}) we find that $\hat{p}_-$ satisfies the 
inhomogeneous Helmholtz equation
\begin{align}
 (\nabla^2  + k^2_-) \hat{p}_-  &= \beta k^2 _-\mathcal{P},
\label{in_helmholtz}
\end{align}
where $k_- = \omega_- /c_0$ is the difference-frequency wavenumber and 
\begin{equation}
 \mathcal{P} = \hat{p}^{*}_{i,1}\hat{p}_{i,2} + \hat{p}^{*}_{i,1}\hat{p}_{s,2} 
+\hat{p}^{*}_{s,1}\hat{p}_{i,2} + \hat{p}^{*}_{s,1}\hat{p}_{s,2},
\label{source}
\end{equation}
with the symbol $^*$ meaning complex conjugation.
The source term $\mathcal{P}$ corresponds to all possible interactions between the primary waves 
which generate the difference-frequency pressure.

\subsection{Boundary conditions}
The uniqueness of solutions of Eqs.~(\ref{helm1}) and (\ref{in_helmholtz})
depend on the acoustic boundary conditions across the scatterer object boundary.
To find these conditions the physical constraints of the scattering problem should be analyzed.

First of all, the presence of primary and secondary pressures induces the object itself to move.
Consequently, an acoustic emission by the object takes place in the host fluid, which means further scattering.
If both the object density is large and its compressibility is small compared to those of the host fluid,
the acoustic emission represents only a small correction to the main scattering due to the presence of the object
in the host fluid~\cite{landau:297}.
In our analysis, this correction is neglected and the object is considered immovable.
Therefore, the boundary condition for a rigid and immovable sphere of radius $a$ is that the normal component of the 
particle velocity should vanish on the sphere's surface.

The particle velocity given up to  second-order approximation is expressed as
\begin{equation}
 {\bf v}  = \varepsilon {\bf v}^{(1)} + \varepsilon^2 {\bf v}^{(2)}, \quad \varepsilon \ll 1,
\end{equation}
where ${\bf v}^{(1)}$ and ${\bf v}^{(2)}$ are the linear and the second-order velocity fields, respectively. 
Thus, 
for the linear velocity we have $\mathbf{v}^{(1)}\cdot \mathbf{e}_r|_{r=a}=0$, where $\mathbf{e}_r$ is the outward normal 
unit-vector on the sphere's surface.
From the linear momentum conservation equation $\rho_0 (\partial \mathbf{v}^{(1)}/\partial t)= - \nabla p^{(1)}$,
we find the following condition for the primary total pressure
\begin{equation}
\label{bc}
\left[\frac{\partial (\hat{p}_{i,n}+\hat{p}_{s,n})}{\partial r}\right]_{r=a} = 0.
\end{equation}
This is known as the Neumann boundary condition.
After substituting Eqs.~(\ref{p_i}) and (\ref{scattered}) into this equation, one obtains the scattering coefficient 
as $s_{nl}^{m} = s_{nl}  a_{nl}^{m}$, where
\begin{equation}
 \label{snl} 
 s_{nl} = - \frac{j'_l(k_n a)}{{h^{(1)}_l}'(k_n a)},
\end{equation}
with the prime symbol meaning derivation.

The second-order particle velocity satisfies the conservation equation~\cite{tjotta:1425}
\begin{equation}
\label{velocity}
\rho_0 \frac{\partial {\bf v}^{(2)}}{\partial t} + \nabla \left(p^{(2)} + \mathcal{L} \right) =0,
\end{equation}
where $\mathcal{L} = (\rho_0/4) \square^2 \phi^{(1)2}$
is the Lagrangian density of the wave, with $ \square^2$ being the d'Alembertian operator.
The function $\phi^{(1)}$ is the first-order velocity potential.
Projecting Eq.~(\ref{velocity}) onto $\mathbf{e}_r$
at the sphere's surface, 
one finds
\begin{equation}
\label{velocity2}
\frac{\partial p^{(2)}}{\partial r}\biggr|_{r=a} = - \frac{\partial \mathcal{L} }{\partial r}\biggr|_{r=a}.
\end{equation}
Now, using the linear 
relation $p^{(1)}=\rho_0(\partial \phi^{(1)}/\partial t)$, Eqs.~(\ref{pmod}) and (\ref{velocity2}),
one obtains the boundary condition for the difference-frequency pressure as
\begin{equation}
\label{velocity3}
\frac{\partial \hat{p}_-}{\partial r}\biggr|_{r=a} = - \frac{k_-^2}{2 k_1k_2}
\frac{\partial \mathcal{P}}{\partial r}\biggr|_{r=a}.
\end{equation}

\subsection{Green's function approach}
The solution of Eq.~(\ref{in_helmholtz}) can be obtained through the Green's function method.
Because the normal derivative of the difference-frequency pressure is specified on the sphere's
surface, the normal derivative of the Green's function on this surface should vanish
in order to avoid overspecification in the method.
Thereby, the difference-frequency pressure amplitude is given in terms of the 
Green's function 
$G({\bf r}|{\bf r}')$ by~\cite{morse:p321} 
\begin{align}
\nonumber
\hat{p}_-(\mathbf{r}) &= -\beta k^2 _- 
\int_V \mathcal{P}(\mathbf{r}') G(\mathbf{r}|\mathbf{r}') dV'
+ \frac{(\nabla^2 - k_-^2)\mathcal{P} }{4 k_1k_2} \\
& -\frac{k_-^2}{2 k_1k_2} \int_S 
\left(\frac{\partial \mathcal{P}}{\partial r'}\right)_{r'=a} G(\mathbf{r}|\mathbf{r}') dS', 
\label{ps}
\end{align}
where $S$ denotes the sphere's surface and
$V$ is the volume of the spatial region from $S$ to infinity.
Note that Eqs.~(\ref{pmod}) and (\ref{velocity3}) have been used in the derivation of Eq.~(\ref{ps}).
The second term in the right-hand side of Eq.~(\ref{ps}) is related to the 
second term in the right-hand side of Eq.~(\ref{pmod}).

The contribution of the surface integral for two interacting
spherical waves (monopoles) is found to be $k_-^3/(k_1k_2)^2$ in~\ref{app:surface}.
In contrast, it will be shown in Eq.~(\ref{pminus}) that the magnitude of the volume 
integral in Eq.~(\ref{ps}) is proportional to $\beta k_-/(k_1k_2)$.
Thus, the ratio of the volume to the surface integral is $k_-^2/(\beta k_1 k_2)$.
It is convenient to write the primary angular frequencies in a symmetric way as follows
$\omega_1=\omega_0 - \omega_-/2$ and $\omega_2 = \omega_0 + \omega_-/2$,
where $\omega_0$ is the mean frequency.
Now the ratio between the integrals can be expressed as $\beta^{-1} [ (\omega_0/\omega_-)^2 - 1/4]^{-1}$.
Note that $\omega_0/\omega_-$ is the downshift ratio.
If the contribution from the surface integral
is about $0.01$ of that from the volume integral in water, the downshift ratio should be larger than $5$.
Therefore, limiting our analysis to downshift ratios larger than $5$, we can neglect the surface integral
in Eq.~(\ref{ps}).

The volume integral in Eq.~(\ref{ps}) can be split into two regions: $a\le r' < r$ 
(inner source volume) and $r < r'$ 
(outer source volume).
In~\ref{app:volume}, the integral corresponding to the 
outer volume is estimated for two interacting
spherical waves.
The result shows that this integral is $O(r^{-2})$.
It will be demonstrated
that the inner volume integral evaluated in
the farfield $k_- r\gg 1$ is $O(r^{-1})$.
Hence, keeping only $O(r^{-1})$ terms in the difference-frequency scattered pressure,
the contribution of the
outer volume integral can be neglected.

The contribution of the second term in the right-hand side of Eq.~(\ref{ps}), i.e. the term related
to local effects, in the farfield is $O(r^{-2})$ as long as the incident waves behave as $O(r^{-1})$ in the farfield. 
Therefore, in the farfield the difference-frequency pressure amplitude is given by
\begin{equation}
 \hat{p}_-(\mathbf{r}) \simeq  -\beta k^2 _-\int_a^r \int_{\Omega} 
\mathcal{P}(\mathbf{r}') G(\mathbf{r}|\mathbf{r}')r'^2 dr' d\Omega', 
\label{df-final}
\end{equation}
where $d\Omega'$ is the infinitesimal solid angle
and the integration is performed on the surface of the unit-sphere $\Omega$. 

In the region $r'<r$, the Green's function which satisfies the Neumann boundary condition 
on the sphere's surface is given by~\cite{morse:p355}
\begin{equation}
G = i k_- \sum_{l,m}  h_l^{(1)}(k_-r) \chi_l(k_-r')
  Y_l^{m}(\theta,\varphi)Y_l^{m*}(\theta',\varphi') ,
\label{green2}
\end{equation}
where 
\begin{equation}
\chi_l(k_-r')= j_l(k_- r') - \frac{j_l'(k_-a)}{{h_l^{(1)}}'(k_-a)}h_l^{(1)}(k_-r').
\label{chi}
\end{equation}
After using the large argument approximation of
the spherical Hankel function~\cite{abramowitz:10} in Eq.~(\ref{green2}), we find
the Green's function in the farfield as
\begin{equation}
G =  \frac{e^{ik_-r}}{r}\sum_{l,m}   i^{-l} \chi_l(k_-r')
  Y_l^{m}(\theta,\varphi)Y_l^{m*}(\theta',\varphi').
\label{green3}
\end{equation}
Now, substituting this equation into Eq.~(\ref{df-final}) along with Eqs.~(\ref{p_i}) and
(\ref{scattered}),
we obtain the  difference-frequency scattered pressure amplitude in the farfield as
\begin{equation}
\label{pminus}
 \hat{p}_-(r,\theta,\varphi) = \frac{\beta k_-}{k_1 k_2 }f_-(r,\theta,\varphi)  \frac{e^{i k_- r }}{r}, \quad k_- r \gg 1,
\end{equation}
where 
\begin{equation}
 f_-(r, \theta,\varphi) = \sum_{l, m}S_l^m(r)Y_l^{m}(\theta,\varphi)
\label{f}
\end{equation}
is the difference-frequency scattering form function.
The interaction function is expressed as	
\begin{align}
\nonumber
S_l^m &= -i^{-l}
\sum_{l_1 ,m_1}\sum_{l_2,m_2} \sqrt{\frac{(2l_1 + 1)(2l_2 + 1)}{4\pi (2l+1)}}\\
\nonumber
&\times C_{l_1, 0, l_2, 0}^{l,0} C_{l_1, m_1, l_2, m_2}^{l, m} 
 a_{1,l_1}^{m_1*}a_{2,l_2}^{m_2}\\
&\times
\biggl(  \varrho_{l_1l_2l}^{(\text{II})} + 
s_{1,l_1}^*\varrho_{l_1l_2l}^{(\text{SI})} 
+ s_{2,l_2}\varrho_{l_1l_2l}^{(\text{IS})} +  s_{1,l_1}^*s_{2,l_2} \varrho_{l_1l_2l}^{(\text{SS})}\biggr),
\label{sl}
\end{align}
where $C_{l_1, m_1, l_2, m_2}^{l, m}$ is the Clebsch-Gordan coefficient, 
which come from the angular integration through the identity~\cite{weisstein:3j_wigner}
\begin{align}
\nonumber
 \int_{\Omega} Y_{l_1}^{m_1} Y_{l_2}^{m_2} Y_{l}^{m} d \Omega &= (-1)^{m} 
\sqrt{\frac{(2l_1 + 1)(2l_2 + 1)}{4\pi (2l+1)}} \\ 
  & \times C_{l_1,0,l_2,0}^{l,0} C_{l_1, m_1, l_2, m_2}^{l,-m}.
\end{align}
The Clebsh-Gordan coefficient satisfies the following conditions~\cite{devanathan:book}
\begin{align}
\nonumber
m_1+m_2&=m,\\
|l_2-l_1|&\le l \le l_1+l_2,
\label{CG_conditions}
\end{align}
otherwise it values zero.
Furthermore, when $m_1=m_2=m=0$ the $l_1+l_2+l$ should be even else the coefficient becomes zero.
The cumulative radial functions $\varrho^{(\cdot\cdot)}$ stands for each possible interaction of the primary waves, i.e.
incident-with-incident (II), scattered-with-incident (SI), incident-with-scattered (IS), and scattered-with-scattered (SS).
They are given by
\begin{align}
 \label{ii}
\varrho_{l_1l_2l}^{(\text{II})} &= k_1k_2k_- \int_a^r  \chi_l(k_- r') j_{l_1}(k_1 r') j_{l_2}(k_2 r') r'^2 dr',\\
\label{is}
\varrho_{l_1l_2l}^{(\text{IS})} &=  k_1k_2k_- \int_a^r  \chi_l(k_- r') j_{l_1}(k_1 r') h_{l_2}^{(1)}(k_2 r') r'^2 dr',\\
\label{si}
\varrho_{l_1l_2l}^{(\text{SI})} &= k_1k_2k_- \int_a^r  \chi_l(k_- r') h_{l_1}^{(2)}(k_1 r') j_{l_2}(k_2 r') r'^2 dr',\\
\label{ss}
\varrho_{l_1l_2l}^{(\text{SS})} &= k_1k_2k_- \int_a^r  \chi_l(k_- r') h_{l_1}^{(2)}(k_1 r') h_{l_2}^{(1)}(k_2 r') r'^2 dr'.
\end{align}

Equations~(\ref{pminus}) and (\ref{f}) along with Eqs.~(\ref{ii})-(\ref{ss}) describe the
difference-frequency generation in the nonlinear scattering of two primary incident waves with arbitrary wavefront 
from a spherical target. 

In the upcoming analysis, it is useful to decompose the difference-frequency pressure amplitude following
the contribution of each primary interaction as given in Eq.~(\ref{sl}).
Accordingly, we  write
\begin{equation}
\label{decomp}
 \hat{p}_- = \hat{p}_-^\text{(II)} + \hat{p}_-^\text{(IS,SI)} + \hat{p}_-^\text{(SS)},
\end{equation}
where the super-indexes stand for the interaction of the primary waves and 
$\hat{p}_-^\text{(IS,SI)} = \hat{p}_-^\text{(IS)}+\hat{p}_-^\text{(SI)}$.
According to Eq.~(\ref{source}) each term in Eq.~(\ref{decomp}) is related
to the primary pressure as follows: $\hat{p}^{*}_{i,1}\hat{p}_{i,2}\rightarrow 
\hat{p}_-^\text{(II)}$, $(\hat{p}^{*}_{i,1}\hat{p}_{s,2} 
+\hat{p}^{*}_{s,1}\hat{p}_{i,2})\rightarrow \hat{p}_-^\text{(IS,SI)}$, and 
$\hat{p}^{*}_{s,1}\hat{p}_{s,2}\rightarrow \hat{p}_-^\text{(SS)}$.

We will show later that the scattered-with-scattered interaction provides the most relevant contribution to 
difference frequency generation analyzed here.
Thus, let us examine the asymptotic behavior of $\varrho_{l_1l_2l}(r)^\text{(SS)}$
with $k_-r\gg 1$.
In doing so, we introduce a new variable $u=r'/r$ in Eq.~(\ref{ss}) and then
\begin{equation}
\varrho_{l_1l_2 l}^\text{(SS)}(r) = k_1k_2k_- r^3 \int_{a/r}^1  
\chi_l(k_- r u) h_{l_1}^{(2)}(k_1 r u) h_{l_2}^{(1)}(k_2 r u) u^2 du.
\end{equation}
Since the integrand uniformly approaches to the product of the 
asymptotic formulas of the spherical functions with large argument
in the interval $a/r\le u\le 1$, then in the farfield this integral has can be written~\cite{bender:249}
\begin{equation}
\varrho_{l_1l_2 l}^\text{(SS)}(r) =  i^{l_1-l_2} \int_{a/r}^1  
\biggl[ \sin\left(k_- r u - \frac{l\pi}{2}\right)- \frac{i^{-l-1} j_l'(k_-a)}{{h_l^{(1)}}'(k_-a)}e^{ik_-ru} \biggr] 
\frac{e^{ik_-r u}}{u} du.
\end{equation}
Therefore,
\begin{eqnarray}
\nonumber
\varrho_{l_1l_2 l}^\text{(SS)}(r) &=&  -\frac{i^{l+ l_1-l_2 - 1}}{2} \biggl\{  \ln\left( \frac{r}{a}\right) - (-1)^l
\left(\frac{2 j_l'(k_-a)}{{h_l^{(1)}}'(k_-a)} - 1 \right)\\
&\times& [i \pi - \text{Ei}(2ik_- a)] \biggr\}, \quad k_- r \gg 1.
\label{rss}
\end{eqnarray}
As a result, the contribution provided by the scattered-with-scattered interaction 
to the difference-frequency scattered pressure varies with the radial distance $r$ as follows
\begin{equation}
\label{pss2}
 \hat{p}_-^\text{(SS)} = A_1 \frac{\ln r}{r} + \frac{A_2}{r},
\end{equation}
where $A_1$ and $A_2$ are constants to be determined from Eqs.~(\ref{pminus})- (\ref{sl}) and (\ref{rss}).
The  $r^{-1}\ln r$ term happens only in regions containing primary energy (volume sources).
Furthermore, it is known as ``continuously pumped sound waves", while the $1/r$ term is called ``scattered sound 
wave''~\cite{berntsen:1968}.

\subsection{Difference-frequency scattered power}
The power scattered at difference-frequency is given by
\begin{equation}
\label{power}
 P_-(r) = \frac{\varepsilon^4 \rho_0 c_0^3 r^2}{2}  \int_{\Omega} \text{Re}\left\{\hat{p}_-^* 
\hat{\bf v}_-\right\}\cdot {\bf e}_r d\Omega,
\end{equation}
where `Re' means the real-part and
the amplitude $\hat{\bf v}_-$ comes from the difference-frequency particle velocity 
${\bf v}_-=\varepsilon^2 c_0 \hat{\bf v}_- e^{-i \omega_- t}$.
In the farfield, cumulative effects are 
dominant in difference-frequency generation.
Thus, referring to Eq.~(\ref{velocity}) we find that $\hat{\bf v}_- \simeq -(i/k_-) \nabla \hat{p}_-$.
After using this result along with Eq.~(\ref{pminus}) and (\ref{f}) into Eq.~(\ref{power}), one obtains
\begin{equation}
\label{power2}
 P_-(r) =  \frac{ \varepsilon^4 \rho_0 c_0^3 \beta^2 k_-^2}{2 k_1^2 k_2^2} \sum_{l,m} \left| S_l^m(r)\right|^2.
\end{equation}

According to Eq.~(\ref{pss2}) the difference-frequency scattered pressurevaries logarithmically with the radial distance $r$.
This result is also known for two concentric outgoing spherical waves~\cite{dean:1039}.
Consequently, the scattered power given in Eq.~(\ref{power2}) will increase without limit as $r\rightarrow \infty$,
unless some account is taken to absorption processes of the primary waves.

\subsection{Series truncation}
To compute Eq.~(\ref{f}), 
we have to estimate \emph{a priori} the number of terms $L_-$ in order to truncate the infinite series.
This is done by performing a truncation of the incident partial-wave expansion given
in Eq.~(\ref{p_i}).
Let  $L_1$ and $L_2$ be the truncation orders corresponding
to the series expansions of the primary waves (incident or scattered) with frequency $\omega_1$ and $\omega_2$,
respectively. 
The parameters $L_1$ and $L_2$ are related, respectively, to the indexes $l_1$ and $l_2$  in Eq.~(\ref{sl}).
To determined $L_1$ and $L_2$, we employ the following rule~\cite{wiscombe:1505,chew:311}
\begin{equation}
 L_n \sim k_n x + c(k_nx)^{1/3}, \quad n=1,2,
\label{rule_Ln}
\end{equation}
where $c$ is a positive constant related to the truncation numerical precision,
and $x$ is a characteristic dimension involved in the wave propagation.
For instance, $x$ can be the scatterer radius or a linear dimension of an interaction
region of the incident waves.
Once $L_1$ and $L_2$ are established,
the truncation order $L_-$ of Eq.~(\ref{f}) is given through Eq.~(\ref{CG_conditions})
as $L_-=L_1+L_2$.

\section{Results and discussion}
\label{results}
\begin{figure}
\centering
\includegraphics[width=.35\textwidth]{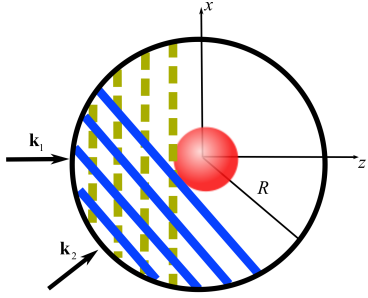}
\caption{
\label{fig:xz} (Color online) Scattering of two confined plane waves from a sphere of radius $a$. 
The wavevectors are denoted by ${\bf k}_1$ and ${\bf k}_2$. 
The incident plane waves propagate only within the spherical region of radius $R$.}
\end{figure}

To illustrate the solution obtained for the difference-frequency scattered pressure given in Eq.~(\ref{pminus}),
we consider a spherical scatterer suspended in water, for which 
$c_0=\unit[1500]{m/s}$, $\rho_0=\unit[1000]{kg/m^3}$, and $\beta = 3.5$ (at room temperature).
The sphere is insonified by two intersecting plane waves which
are confined in a spherical region of radius $R$.
This region is centered on the scatterer as shown in Fig.~\ref{fig:xz}.
The incident wavevectors are denoted by  ${\bf k}_1$ and ${\bf k}_2$.
Yet this model is not entirely realistic, spatially confined plane waves with fast spatial decay
can be experimentally produced by means of focused transducers~\cite{chen:313}.

The partial wave expansion of each plane wave is given by~\cite{colton:3}
\begin{equation}
 \hat{p}_{i,n}  =  4\pi \sum_{l,m}i^lY_l^{m*}(\theta_n,\varphi_n)j_l(k_nr)
Y_l^m(\theta,\varphi),\quad r\le R,
\label{arb_plane}
\end{equation}
where $n=1,2$ and ${\bf k}_n$
is given in terms of $(k_n,\theta_n,\varphi_n)$,
with $\theta_n$ and $\varphi_n$ being the polar and azimuthal angles, respectively.
Comparing Eqs.~(\ref{arb_plane}) and (\ref{p_i}) we find that the beam-shape coefficient is given by
\begin{equation}
 a_{nl}^m =  4\pi i^lY_l^{m*}(\theta_n,\varphi_n).
\end{equation}
For radial distances larger than $R$ the incident pressure amplitude vanishes, 
i.e. $\hat{p}_{i,n}=0$.
Hence, the integration interval of Eqs.~(\ref{ii})-(\ref{si}) should be $a\le r'\le R$.

The scattering problem can be further simplified by assuming that one plane wave propagates along the $z$-axis, thus,  
$\mathbf{k}_1 = k_1 \mathbf{e}_z$, 
with ${\bf e}_z$ is the Cartesian unit-vector along
the $z$-axis.
Whereas the other wave travels  along the direction determined by 
$\mathbf{k}_2 =  \sin (\theta_2){\bf e}_x + k_2 \cos(\theta_2) {\bf e}_z$, where
${\bf e}_x$ is the Cartesian unit-vector along
the $x$-axis.
Unless specified, the scatterer radius is $a=\unit[1]{mm}$, the 
the cspherical region of the plane waves has radius $R=\unit[2.4]{mm}$,  
the radial observation distance is $r=\unit[0.1]{m}$, and
the mean- and the difference-frequency are 
$\omega_0/2\pi = \unit[1.5]{MHz}$ and $\omega_-/2\pi=\unit[100]{kHz}$, respectively.
Thus, the downshift ratio is fifteen.
These parameters are in the same range as those used in some nonlinear
acoustical imaging methods~\cite{sato:49,kim:175,chen:313}.  
The size factors involved in the scattering problem are 
$k_- R = 1$, $k_1R = 14.5$, $k_2R = 15.5$, $k_- a= 0.41$, $k_1 a=6.07$, $k_2a=6.49$, $k_-r=41.88$,
$k_1r =607.37$, and $k_2r=649.26$.
The truncation orders are determined by
setting the parameter $c=4$ (four precision digits) in Eq.~(\ref{rule_Ln}).
Hence, the truncation orders for $\hat{p}_-^\text{(II)}$, $\hat{p}_-^\text{(IS)}$,
 $\hat{p}_-^\text{(SI)}$, and $\hat{p}_-^\text{(SS)}$ are respectively given by
$(L_-,L_1,L_2)=(69,33,36), (48,33,15),(51,15,36), (29,14,15).$

The integrals in Eqs.~(\ref{ii})-(\ref{ss}) can be solved analytically for arbitrary 
combinations of the indexes $l,l_1,$ and $l_2$.
Nevertheless, the number of terms in the solution grows combinatorially with the indexes.
In the present example, the analytic solution of the integrals seems not to be practical.
Hence, the integrals are solved numerically by using the Gauss-Kronrod quadrature method~\cite{matlab:gk}.

\begin{figure*}
\centering
\subfigure[]{\includegraphics[width=.49\textwidth]{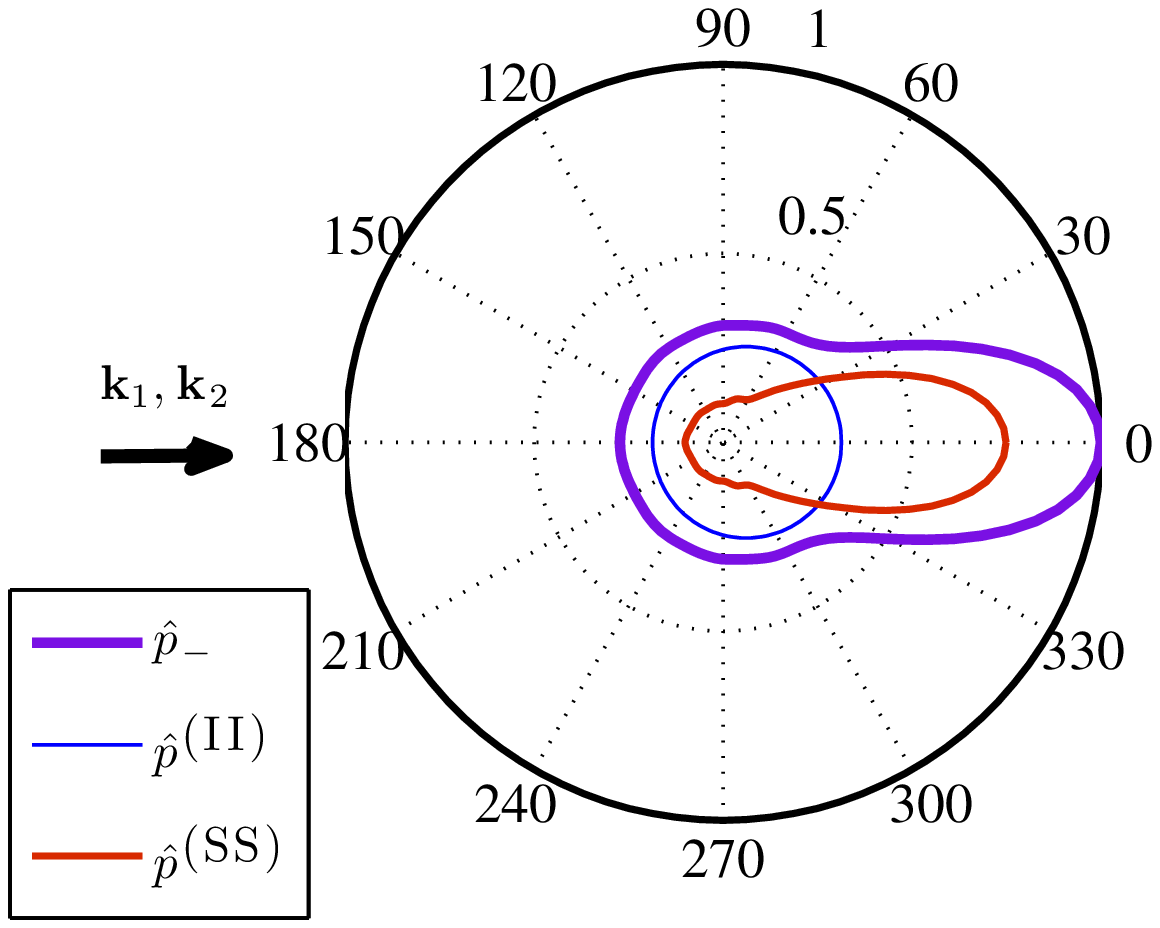}}
\subfigure[]{\includegraphics[width=.49\textwidth]{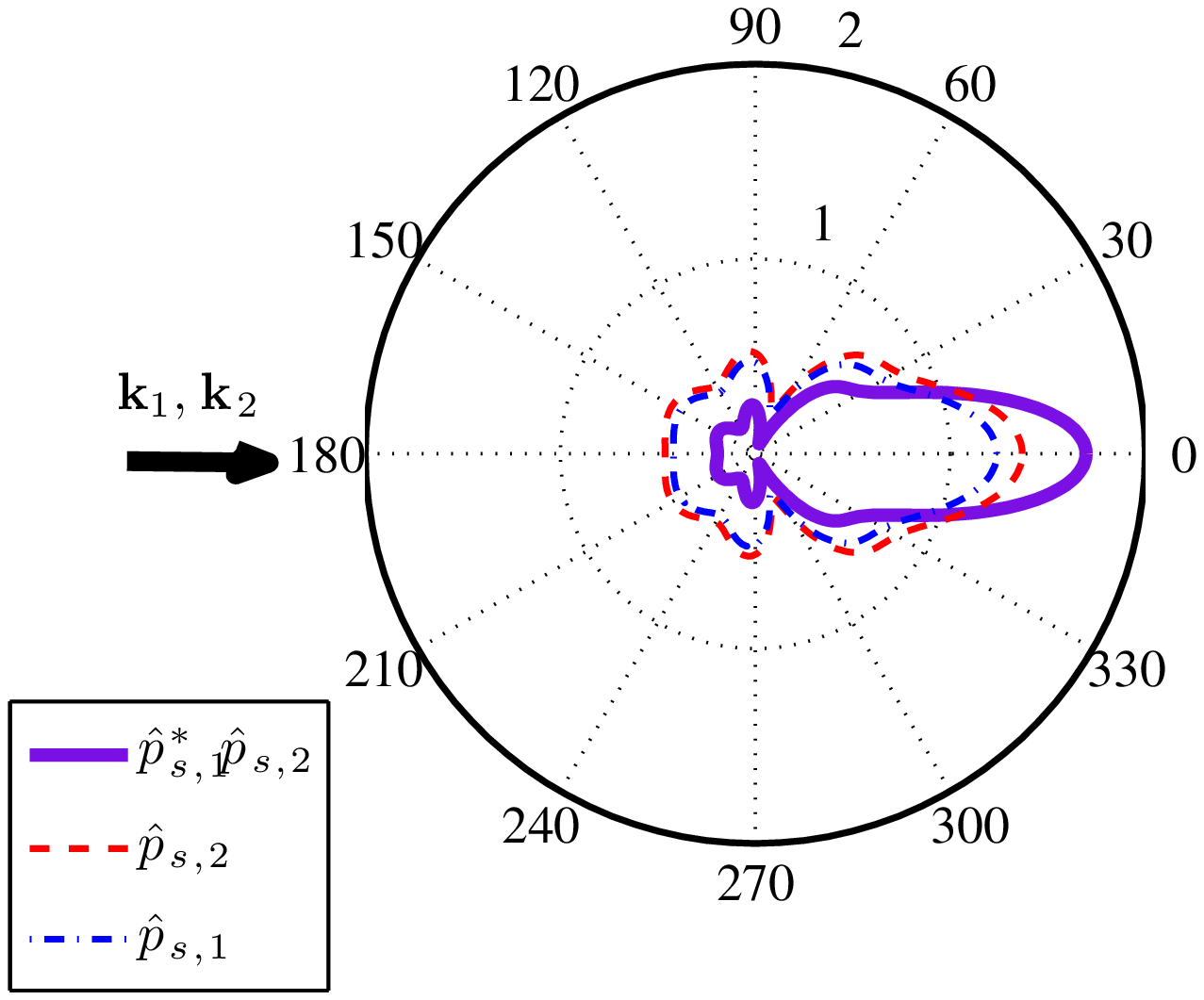}}
\caption{
\label{fig:p_all0}
(Color online)  The directive pattern in the $xz$-plane of (a)
the difference-frequency scattered pressure (normalized to maximum value of $|\hat{p}_-|$ which is $0.0823$) 
generated by two collinear plane waves, 
and (b) the linear scattered pressures.
The physical parameters used here are
$r=\unit[0.1]{m}$, $R=\unit[2.4]{mm}$, $a=\unit[1]{mm}$, $\omega_0/2\pi = \unit[1.5]{MHz}$, and 
$\omega_-/2\pi=\unit[100]{kHz}$.
The corresponding size factors are 
$k_- R = 1$, $k_1R = 14.5$, $k_2R = 15.5$, $k_- a= 0.41$, $k_1 a=6.07$, $k_2a=6.49$, $k_-r=41.88$,
$k_1r =607.37$, and $k_2r=649.26$.
The arrows indicate the direction of the incident wavevectors.}
\end{figure*}

The directive pattern in the $xz$-plane of the difference-frequency 
scattered pressure given in Eq.~(\ref{pminus}) and produced by two collinear plane waves
($\theta_2 = 0$) are shown in Fig.~\ref{fig:p_all0}.
The dimensionless pressures $\hat{p}_-^\text{(II)}$, $\hat{p}_-^\text{(IS,SI)}$,  and $\hat{p}_-^\text{(SS)}$
are also exhibited.
The magnitudes of these functions  are normalized to the maximum value of 
$|\hat{p}|$  which is $0.0807$.
The contribution from $\hat{p}_-^\text{(IS,SI)}$ is small compared to other dimensionless pressures.
In the region $30^\circ <\theta<330^\circ$,
the difference-frequency scattered pressure is dominated by
$\hat{p}_-^\text{(II)}$.
Both $\hat{p}_-^\text{(II)}$ and $\hat{p}_-^\text{(SS)}$ give a prominent contribution to the difference-frequency
scattered pressure when 
$\theta<30^\circ$ and $\theta>330^\circ$.
In this case, the contribution of $\hat{p}_-^\text{(II)}$ corresponds to $25\%$ of the scattered difference-frequency pressure.
The magnitude of this pressure mostly occurs in the forward scattering direction $(\theta = 0^\circ)$.
We notice that as the radius $R$ of the spherical region increases, the role of $\hat{p}_-^\text{(II)}$ overcomes 
the contribution of the scattered-with-scattered interaction.
The spatial behavior of  $\hat{p}_-^\text{(SS)}$  resembles that of the linear scattered pressure by the sphere as 
shown in Fig~\ref{fig:p_all0}.b.

The dimensionless pressure $\hat{p}_-^\text{(II)}$ is related 
to a parametric array whose primary waves are confined in the spherical region of radius $R$.
We can obtain an approximate solution of the parametric array pressure in the farfield, when
${\bf r} = r {\bf e}_z$ (forward scattering direction $\theta = 0^\circ$).
To calculate the parametric array pressure
we consider the source term in Eq.~(\ref{df-final}) as $\mathcal{P} =  e^{i k_- r' \cos
\theta'}$.
Moreover, we approximate 
the Green's function in the farfield to
\begin{equation}
G = \frac{i k_- (r-r'\cos\theta')}{4 \pi r}.
\end{equation}
Thus, substituting the source term and
the Green's function into Eq.~(\ref{df-final}),
we find that the dimensionless parametric array pressure 
is given by
\begin{equation}
 \hat{p}_-^\text{(PA)} \simeq - \beta k_-^2R^3 \frac{e^{i k_- r}}{3 r}, \quad \theta = 0^\circ.
 \label{ii-parametric}
\end{equation}
Using the physical parameters of Fig.~\ref{fig:p_all0}, we find good agreement between the
this pressure and  $\hat{p}_-^{\text{(II)}}$, with relative error  smaller than $9\%$.
This error might be caused among other things by the presence of the scatterer in the spherical confining region,
which is not accounted by Eq.~(\ref{ii-parametric}).

\begin{figure*}
\subfigure[]{\includegraphics[width=.5\textwidth]{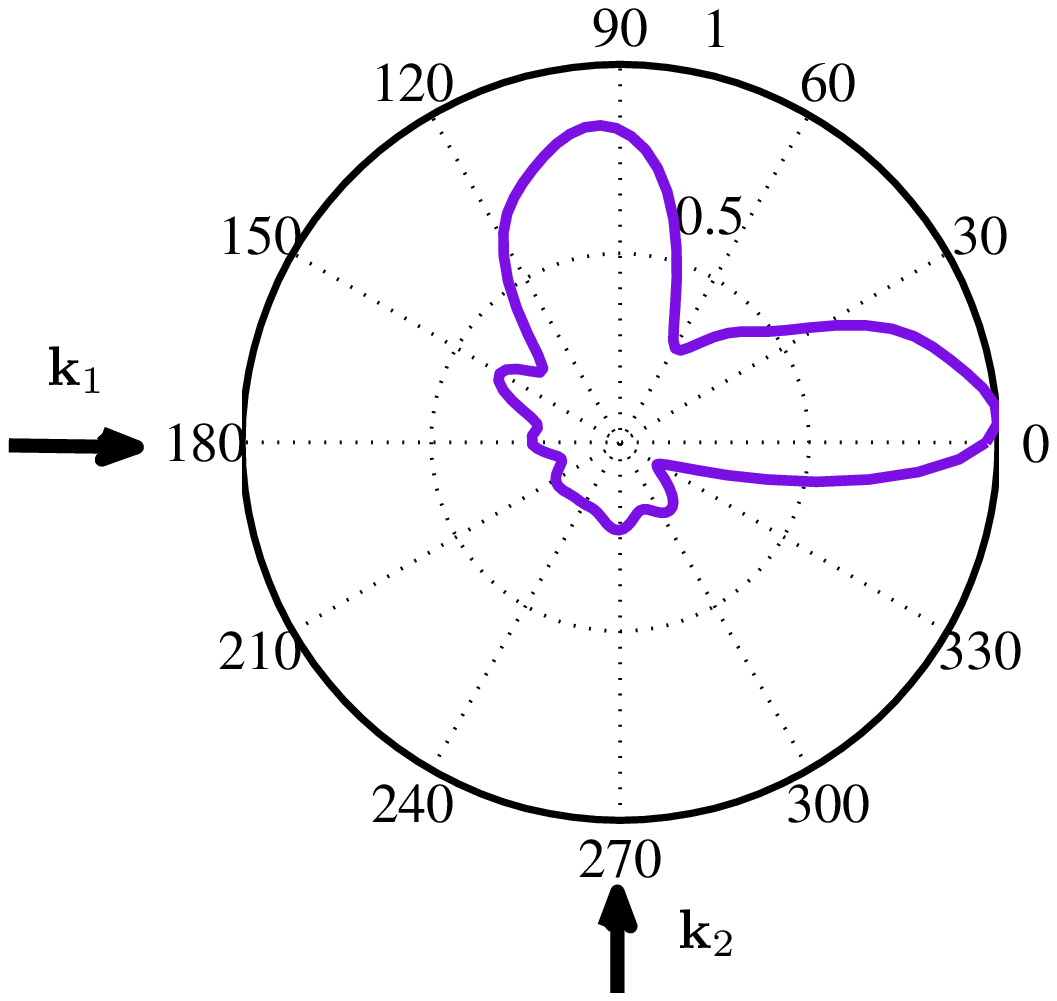}}
\subfigure[]{\includegraphics[width=.5\textwidth]{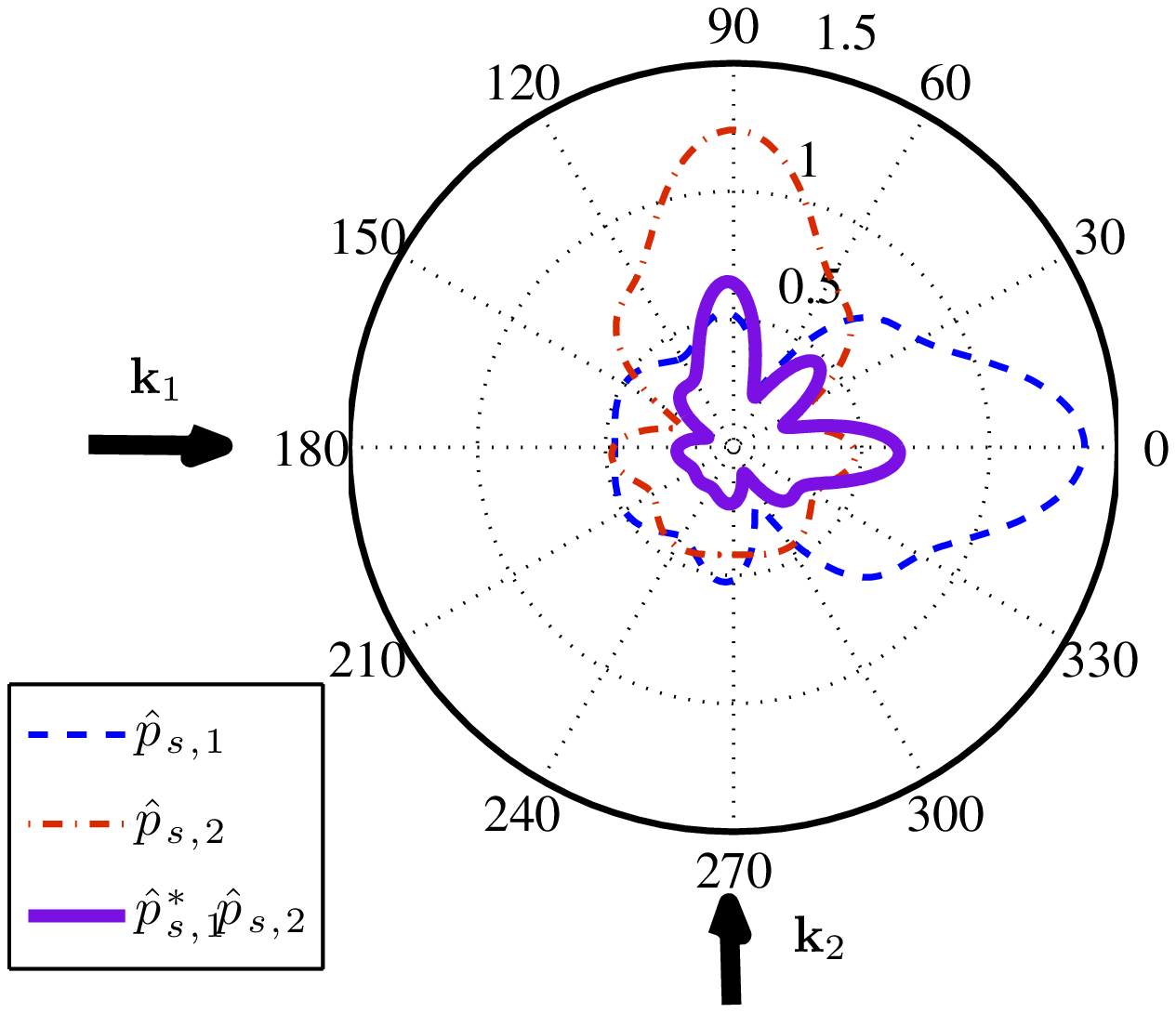}}
\caption{
\label{fig:p_all90}
The directive pattern in the $xz$-plane of (a) the difference-frequency pressure (normalized to $0.0134$) produced by 
two perpendicular 
plane waves, and (b) the linear scattered pressures.
The physical parameters used in the evaluation here
are the same as those described in Fig.~\ref{fig:p_all0}.
The arrows point to the direction of the incident wavevectors.
}
\end{figure*}

The directive pattern in the $xz$-plane of  the difference-frequency scattered pressure
produced by two intersecting plane waves at a
right angle $(\theta_2=90^\circ)$ is displayed in Fig.~\ref{fig:p_all90}.
The component $\hat{p}_-^\text{(II)}$ corresponds
to less $1\%$ of the total pressure and it cannot be
seen in this figure.
This result is in agreement with early studies which state that two intersecting plane waves
at right angle do not produce difference-frequency pressure outside
the intersecting region~\cite{westervelt:199}.
The term $\hat{p}_-^\text{(IS,SI)}$ does not contribute significantly to difference-frequency scattered pressure.
Thus, $\hat{p}_-^\text{(SS)}$ is responsible for this pressure.
The two mainlobes of the difference-frequency scattered pressure lies on the forward scattering 
directions $(\theta=0^\circ, 90^\circ)$  of each incident wave as depicted 
in Fig.~\ref{fig:p_all90}.b.
Furthermore, these lobes follow the pattern of the linear scattered mainlobes
as shown in~\ref{fig:p_all90}.b. 

\begin{figure*}
\subfigure[]{
\includegraphics[width=.5\textwidth]{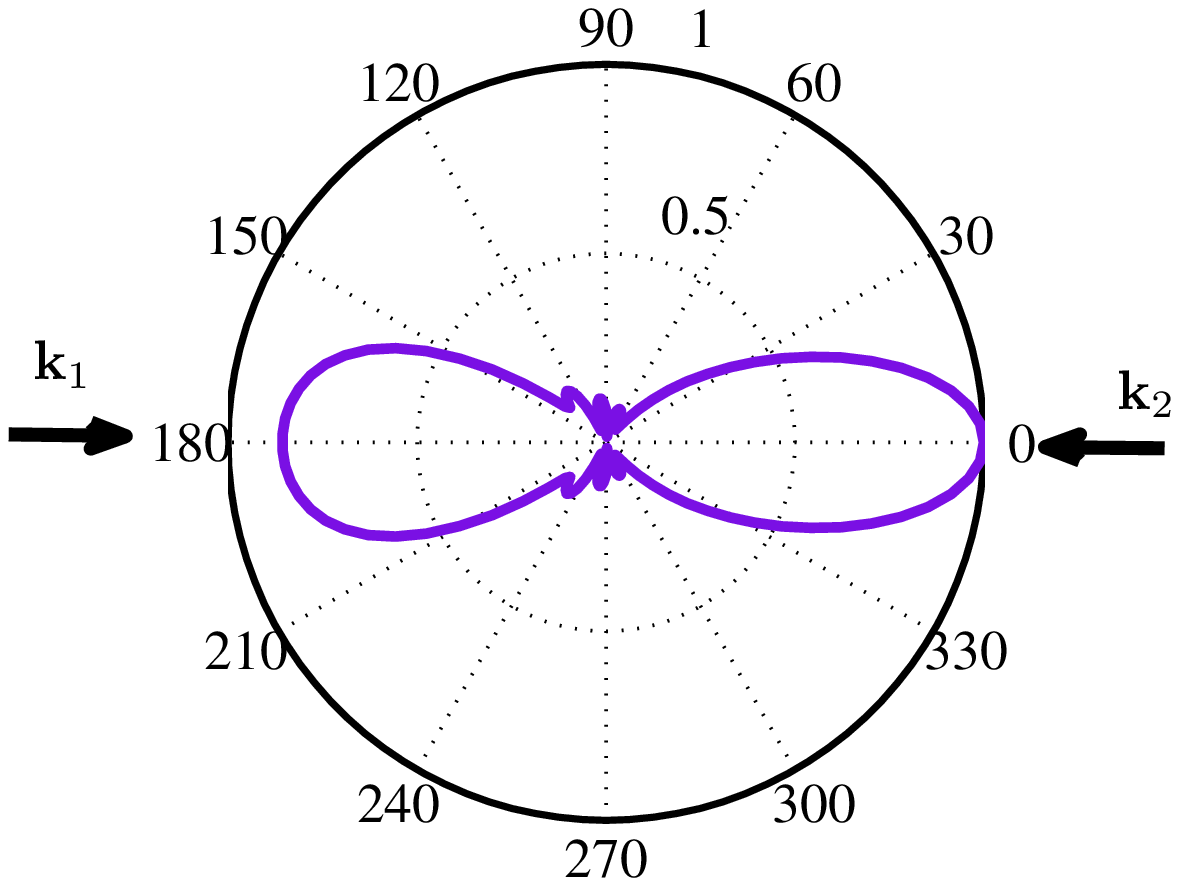}}
\subfigure[]{
\includegraphics[width=.5\textwidth]{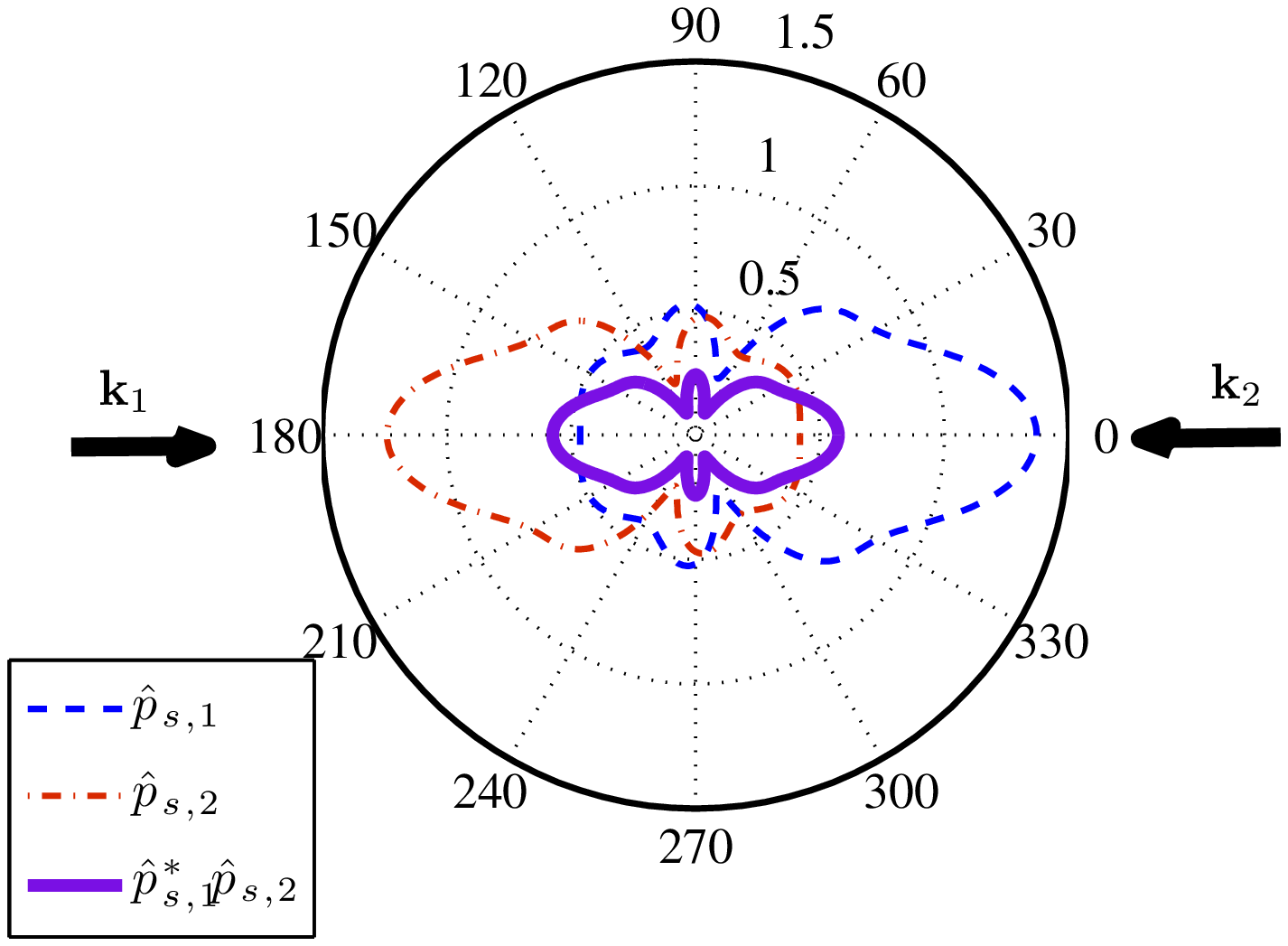}}
\caption{
\label{fig:p_all180}
The directive pattern  in the $xz$-plane of (a) the difference-frequency scattered pressure (normalized to $0.0178$) due to 
two counter-propagating  plane waves, and (b) the linear scattered pressures.
The physical parameters used here
are the same as those described in Fig.~\ref{fig:p_all0}.
The arrows point to the direction of the incident wavevectors.}
\end{figure*}

In Fig.~\ref{fig:p_all180}, we show
the directive pattern in the $xz$-plane of the difference-frequency
scattered pressure generated in the scattering of two 
counter-propagating plane waves $(\theta_2=180^\circ)$.
The contributions of $\hat{p}_-^\text{(II)}$ and $\hat{p}_-^\text{(IS,SI)}$ are small compared to that 
from $\hat{p}_-^\text{(SS)}$.
It is known that the counter-propagating  waves weakly 
interact nonlinearly~\cite{gusev:211}.
Thus, the difference-frequency pressure is practically due to $\hat{p}_-^\text{(SS)}$.
The pressure is not symmetric due to a difference in the incident wave frequencies.
The difference-frequency pressure follows the behavior of the linear scattered pressures shown in 
Fig.~\ref{fig:p_all180}.b.

\begin{figure}[h]
\centering
\includegraphics[width=.65\textwidth]{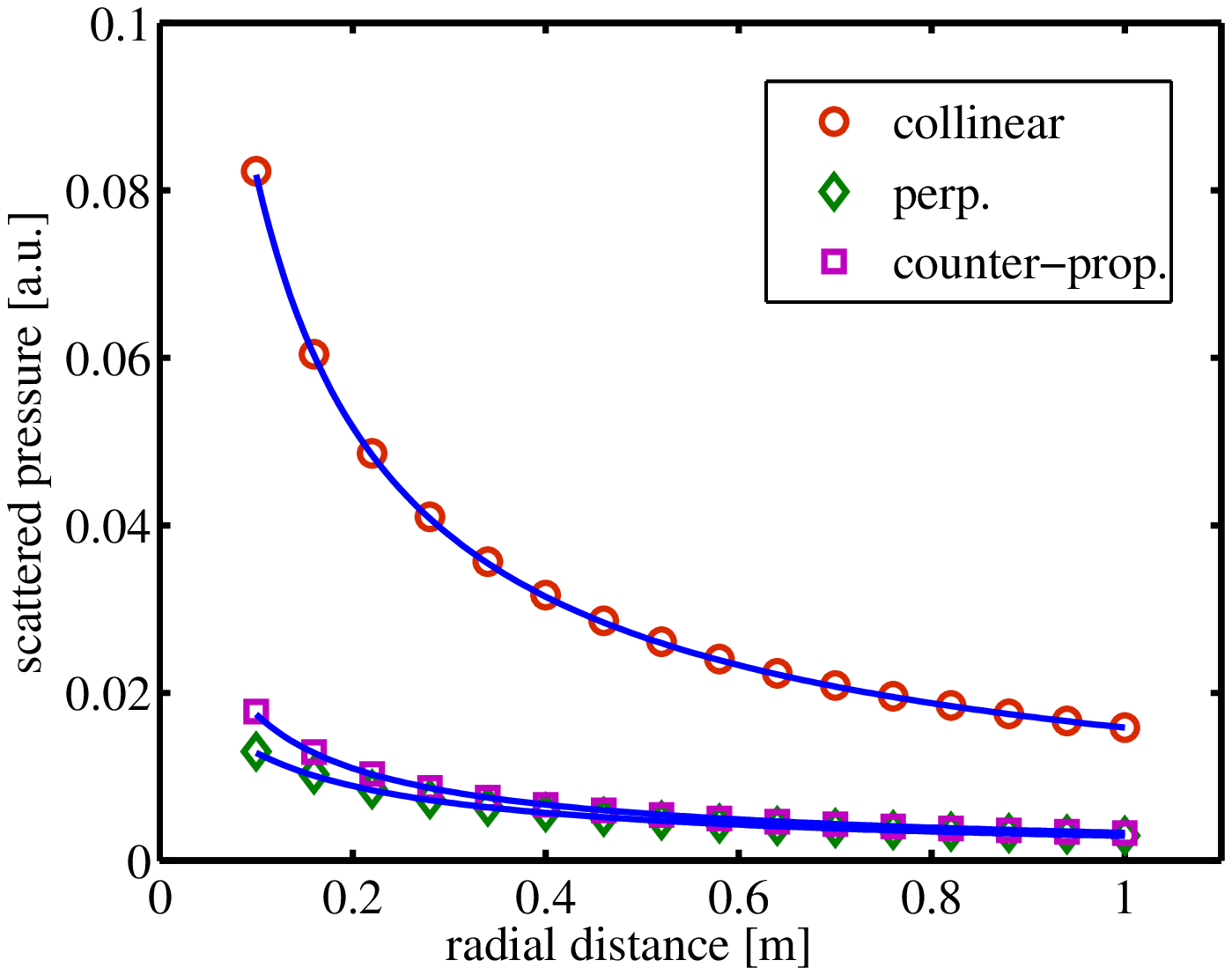}
\caption{
\label{fig:p_dist}
The (dimensionless) difference-frequency  pressure magnitude in the forward scattering direction $\theta=0^\circ$ 
varying with the radial distance $r$.
The physical parameters used in here
are the same as those described in Fig.~\ref{fig:p_all0}.}
\end{figure}

The dependence of the difference-frequency scattered pressure with the radial distance $r$ is exhibited
in Fig.~\ref{fig:p_dist}.
The pressure is calculated in the forward scattering direction $\theta=0^\circ$.
In all cases, the main contribution to this pressure comes from $\hat{p}_-^\text{(SS)}$.
analyzed here.
Note that according to Eq.~(\ref{pss2}), the difference-frequency scattered pressure 
varies as $A_1r^{-1}\ln r + A_2r^{-1}$.

\begin{figure}[h]
\centering
\includegraphics[width=.65\textwidth]{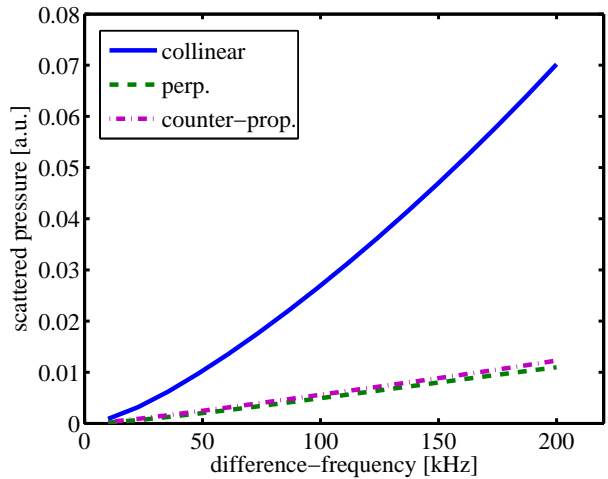}
\caption{
\label{fig:p_df}
The (dimensionless) scattered pressure magnitude versus the difference-frequency.
The physical parameters used in the evaluation here
are the same as those described in Fig.~\ref{fig:p_all0}.
}
\end{figure}

The scattered pressure varying with  difference-frequency is shown in Fig.~\ref{fig:p_df}.
The pressure is evaluated at $r=\unit[0.5]{m}$ in the forward scattering direction $\theta=0^\circ$.
In all configurations, the scattered pressure increases with  difference-frequency.
The difference-frequency scattered pressures due to the perpendicular and  counter-propagating incident plane waves have very close magnitudes. 
According to Eqs.(\ref{pminus}) and (\ref{pss2}), the scattered pressure varies with difference-frequency as $\omega_- f(\omega_-)$, where $f$ is a function determined
in these equations. 
Moreover, by referring to Eq.~(\ref{pminus}) one can show that the difference-frequency scattered pressure diverges when  
$\omega_-\rightarrow 2\omega_0$ and $\omega_1\rightarrow 0$.
Physically the scattered pressure does not diverge, but decays due to attenuation instead. 

It is worthy to relate our analysis  with a previous theoretical study on difference-frequency generation in acoustic
scattering~\cite{abbasov:158}.
We have tried to draw a direct comparison between this work and the method presented here.
Unfortunately, we could not reproduce the reference'ss results due to the presence of angular singularities in the 
difference-frequency scattered fields.
Therefore, no comparison was possible. 
Furthermore, we did try to explain the experimental results of difference-frequency generation in the scattering given in 
Ref.~\cite{abbasov:309}.
In this study, a nonlinear scattering experiment was performed involving two collinear beams and a spherical target.
The incident waves are generated by a circular flat transducer.
Despite the authors claim that the incident beams approach to plane waves,
the directive patterns of the linear scattered waves obtained in the experiments
do not follow this assumption (see Ref.~\cite{pierce:431}).
Since the scattering does not involve incident plane waves, a direct comparison
of our theory (for plane waves) and the experimental results is not reasonable.
However, one of the conclusions of Ref.~~\cite{abbasov:309}  is that 
the difference-frequency scattered pressure is mostly produced
by the incident-with-incident and the scattered-with-scattered interactions.
This conclusion is also supported by our  results.

\section{Summary and conclusions}
The difference-frequency generation in the scattering of two interacting acoustic waves
with an arbitrary wavefront by a rigid sphere was theoretically analyzed.
The difference-frequency scattered pressure in the farfield was obtained as a 
partial-wave series expansion.
The amplitude of each  partial-wave is given
by the interaction function $S_l^m$, which
depends on the observation distance from the scatterer,
the beam-shape and scattering coefficients 
of the primary waves.
The developed method was applied to the scattering of two intersecting plane 
waves located within a spherical region.
The directive pattern of the difference-frequency scattered pressure 
was analyzed in three  incident wave configurations: collinear, perpendicular, and 
counter-propagating.
In the collinear arrangement, the incident-with-incident and  scattered-with-scattered interactions  provide
a more prominent contribution to the scattered pressure.
In all other configurations, the scattered-with-scattered interaction
prevails over the other interactions.
The results show that the scattered pressure increases with difference-frequency.
Experimental evidence of this feature was reported in  Ref.~\cite{silva:5985}.
Moreover, the scattered pressure was shown to vary with the observation distance as $r^{-1}\ln r$ and $1/r$.

Sound absorption effects in the fluid were not considered.
If only compressional waves are assumed to propagate in a weakly viscous fluid,
the proposed model can readily accommodate absorption effects by changing the wavenumber 
$k\rightarrow k + i \alpha$, where $\alpha$ is the absorption coefficient.
Attenuation may affect the obtained results here in at least one way.
Both incident and scattered waves at the fundamental frequencies $\omega_1$ and 
$\omega_2$ are more attenuated than the difference-frequency scattered wave.
Thus, the nonlinear interaction range of the fundamental waves in a viscous fluid
is shorter than in a nonviscous fluid.
Consequently, a less difference-frequency scattered signal is supposed to be formed
in a viscous fluid.

In conclusion, this article presents the difference-frequency generation in nonlinear
acoustic scattering of two incident waves with an arbitrary wavefront.
This study can help unveil important features of acoustic scattering not dealt with
before.

\section*{Acknowledgements}
This work was supported by grants 306697/2010-6 CNPq, 477653/2010-3 CNPq, 2163/2009 PNPD--CAPES,
and FAPEAL (Brazilian agencies).

\appendix
\section{Surface integral}
\label{app:surface}
According to Eq.~(\ref{velocity3}) the surface integral in Eq.~(\ref{ps}) is given by
\begin{equation}
 I_S = - \frac{(k_-a)^2}{2 k_1k_2}\int_{\Omega} 
G( r,\theta,\varphi | a, \theta',\varphi')\frac{\partial \mathcal{P}}{\partial r'}\biggr|_{r'=a} d\Omega'.
\label{IS}
\end{equation}
This integral will be estimated for two interacting spherical waves.
Thus, the source term $\mathcal{P}$ is given by 
\begin{equation}
\mathcal{P}(r') = \frac{e^{ik_-r'}}{k_1 k_2 r'^2}.
\label{PA}
\end{equation}
From Eq.~(\ref{green3}) the Green's function becomes 
\begin{equation}
 G= k_- a \frac{e^{ik_- r}}{r} \sum_{l,m} \frac{(-i)^l}{{h^{(1)}_l}'(k_-a)}Y_l^m(\theta',\varphi')Y_l^m(\theta,\varphi).
\label{GA}
\end{equation}
Substituting Eqs.~(\ref{PA}) and (\ref{GA}) into Eq.~(\ref{IS}), yields
\begin{equation}
I_S = \frac{k_-^3}{2 k_1^2 k_2^2 } (2i+k_- a)
\frac{e^{ik_- (r-a)}}{r}.
\end{equation}

\section{Outer volume integral}
\label{app:volume}
The outer volume integral reads
\begin{equation}
I_\infty = \beta k_-^2 \int_r^\infty \int_{\Omega} G_\infty({\bf r}| {\bf r}')\mathcal{P}({\bf r}')r'^2 dr' d\Omega',
\label{IV}
\end{equation}
where
the Green's function is given by~\cite{morse:p354}
\begin{equation}
G_\infty = i k_- \sum_{l,m}   \chi_l(k_- r)h_l^{(1)}(k_-r') Y_l^{m}(\theta,\varphi) Y_l^{m*}(\theta',\varphi'),
\label{green4}
\end{equation}
with $a\le r<r'$.
We assume that the source term is due the interaction of two spherical waves as given in 
Eq.~(\ref{PA}).
By substituting Eqs.~(\ref{PA}) and (\ref{green4}) into Eq.~(\ref{IV}), one finds
\begin{equation}
I_\infty = \frac{\beta k_-}{k_1 k_2}\chi_0(k_- r) \int_r^\infty \frac{e^{2ik_-r'}}{r'} dr'.
\label{IV2}
\end{equation}
After integrating by parts, 
we obtain
\begin{equation}
I_\infty = \frac{\beta k_-}{k_1 k_2} \chi_0(k_- r)
\left[
\frac{e^{2ik_-r}}{r} + O(r^{-2})\right].
\label{I-infty}
\end{equation}
Therefore, evaluating $\chi_0(k_- r)$ through the expressions of the spherical functions, we find 
$I_\infty = O(r^{-2}) $.

\small

\end{document}